\documentclass[aps,final,notitlepage,oneside,onecolumn,nobibnotes,nofootinbib,superscriptaddress,noshowpacs,centertags]{revtex4-1}
\usepackage[english]{babel}
\usepackage{graphicx}
\usepackage{latexsym}
\usepackage{amssymb}
\usepackage{amsmath}
\usepackage{float}

\begin{document}

\title{Recurrent fast radio bursts from collisions of neutron stars \\ in the evolved stellar clusters}
\author{V. I. Dokuchaev}\thanks{e-mail: dokuchaev@inr.ac.ru}
\affiliation{Institute for Nuclear Research of the Russian Academy of Sciences, Moscow, Russia}
\affiliation{National Research Nuclear University MEPhI (Moscow Engineering Physics Institute), Moscow, Russia}
\author{Yu. N. Eroshenko}\thanks{e-mail: eroshenko@inr.ac.ru}
\affiliation{Institute for Nuclear Research of the Russian Academy of Sciences, Moscow, Russia} 

\date{\today}

\begin{abstract}
We propose the model describing the observed multiple fast radio bursts due to the close encounters and collisions of neutron stars in the central clusters of the evolved galactic nuclei. The subsystem of neutron star cluster may originate in the dense galactic nucleus evolutionary in the combined processes of stellar and dynamical evolution. The neutron stars in the compact cluster can produce the short living binaries with the highly eccentric orbits, and finally collide after several orbital revolutions. Fast radio bursts may be produced during the close periastron approach and at the process of the final binary merging. In the sufficiently dense star cluster the neutron stars collisions can be very frequent. Therefore, this model can explain in principle the observed recurrent (multiple) fast radio bursts, analogous to the observed ones from the source FRB~121102. Among the possible observational signatures of the proposed model may be the registration of the gravitational wave bursts by the laser interferometers LIGO/VIRGO or by the next generation of gravitational wave detectors.
\end{abstract}

\maketitle 


\section{Introduction}

In the recent years the enigmatic fast radio bursts (FRBs) were observed \cite{Lorimer07,Keane12,Keane12,Thornton13,Spitler14,Burke14,Petroff15,Ravi15,Rane16}. Nowadays, the physical nature of these bursts is quite unknown. The numerous models were proposed for explanation of the FRBs. In particular, the bursts in magnetars \cite{PopPos13,Kiyoshi15}, the giant impulses from the young rapidly rotating pulsars \cite{LyuBurPop16}, the gravitational collapses of neutron stars to black holes \cite{FalRez14}, the impact of a supernova shock on the magnetosphere of a neutron star \cite{EgoPos09}, collisions of asteroids and comets with neutron stars \cite{GenHua15}, collisions of axion miniclusters with neutron stars \cite{Iwa15}, \cite{Iwa14} or explosive decay of axionic miniclusters \cite{Tka14}, binary collisions of neutron stars \cite{Tom13,CalKanWei16,DaiWanWu16,Wanetal16}, collapse of the magnetospheres of Kerr-Newman black holes \cite{Liuetal16} and even the activity of extraterrestrial civilizations \cite{LinLoe16} were presented as possible explanations.

Besides the solitary FRBs it was registered the source FRB~121102, generating the multiple recurrent FRBs \cite{Spietal16,Schetal16}. This source was associated with the galaxy at the red-shift $z\approx0.2$ \cite{Shretal16}. This galaxy reveals no activity and the source of FRBs is shifted possibly from the galactic center. Other observations, including the VLA, demonstrate the coincidence of FRB~121102 with the permanent radio-source of the size less then $1$~pc and, is possibly related with the host galactic nuclei of low-activity \cite{Maretal16}, \cite{Chaetal16}. If FRB~121102 doesn't represent some exceptional case, the crucial feature of the FRBs phenomenon is, therefore, their unusual recurrence. The possible model for explanation of the recurrent FRBs is the close encounters of asteroids with the strongly magnetized neutron star \cite{Daietal16} or repeated accretion and bursts in the neutron star-white dwarf binary \cite{Guetal16}. 

Sometimes it was claimed that it is impossible to explain the multiple FRBs in the models with collision events because the source is destroyed and collisions do not repeat. Nervelessness, in this work we propose the model of the recurrent FRBs from the binary collisions of neutron stars in the central clusters of stars in the galactic nuclei. The recurrence is inevitable if collisions take place in the dense central stellar clusters of the galactic nuclei, containing large fraction of neutron stars. In this case it is possible to explain not only the few FRBs events from the same source, but also the prolonged recurrence of FRBs. The coherent radio emission from the collisions of neutron stars was considered in \cite{LipPan96,PshPos10,Lyu13}. 

Neutron star clusters in the galactic nuclei may evolutionary originate in the combined processes of stellar and dynamical evolution \cite{Col67,San70,QuiSha90}. During this evolution the most massive stars settle down close to the cluster center (mass segregation) and finally transform to the neutron stars and black holes. In result the cores of the central clusters in some galactic nuclei consist of neutron stars and black holes. The subsequent random collisions of these star remnants provides the growing accumulation of massive black holes in the galactic nuclei. This process, in particular, may provide the formation of black holes with mass $\sim30M_\odot$, responsible for the generation of the observed gravitational wave signal by the LIGO detector during coalescence of two black holes \cite{LigoVirgo}. Note that in the hierarchic scenario of galaxy formation the dense neutron star cluster may be situated beyond the galactic center due to the galaxies merge. 

The recurrent collisions of neutron stars in the evolved stellar cluster  were considered, in particular, in \cite{DokEroOze98,BerDokHidden,BerDokHidden06,DokEro11}. For generation of the FRBs it is not necessary the direct neutron star collision. Instead of, it is sufficient even  close encounter of magnetized neutron stars providing the interaction of their magnitospheres, accompanied by the generation of FRBs (e.\,g., due to the reconnections of magnetic field lines).

\section{Central stellar clusters in the galactic nuclei}
\label{centrsec}

The star clusters in the galactic nuclei may contain the large fraction of compact stellar remnants in the form of neutron stars and stellar mass black holes, which form the compact central subclusters \cite{Col67,San70,QuiSha90,Feretal06}. Coalescences and collisions of these compact stellar remnants may generate in principle the highly energetic events, in particular, in the form of gravitational waves and FRBs. 

Let us consider the stellar system with a mass $M$ and virial radius $R$, consisting of the mixture of $N\gg 1$ neutron stars and stellar mass black holes. For simplicity and brevity we will call this multicomponent stellar cluster as the ``neutron star cluster''. The velocity dispersion of stars in this cluster is $v\simeq(GM/2R)^{1/2}$. The dynamical evolution of the dense stellar cluster proceeds mainly due to the pair relaxation. The characteristic time of binary relaxation is 
\begin{equation}
 \label{tr}
 t_r=\left(\frac{2}{3}\right)^{\!1/2}\!\!\!
 \frac{v^3}{4\pi G^2m^2n\Lambda},
\end{equation}
where $\Lambda=\ln(0.4N)$ is the Coulomb logarithm, and $m$ is the stellar mass. The stars are ``evaporated'' from the core of the cluster, producing the extensive near isothermal ``atmosphere'' around the central core. The remained core of the cluster is secularly contracting, and the velocity dispersion of stars is growing. 

In the homological approximation the core mass $M_c$ and the core radius $R_c$ of the stellar cluster evolves as
\begin{equation}
M_c\propto[1-(t-t_i)/(\kappa t_{ri})]^{\nu_1}, \quad R_c\propto[1-(t-t_i)/(\kappa
t_{ri})]^{\nu_2}.
\label{mr}
\end{equation}
In this equation ``i'' denotes the initial values, $\nu_1=2\alpha_2/(7\alpha_2-3\alpha_1)$,
$\nu_2=2(2\alpha_2-\alpha_1)/(7\alpha_2-3\alpha_1)$,
$\kappa=2/(7\alpha_2-3\alpha_1)$, where $\alpha_1=8{.}72\times10^{-4}$ and $\alpha_2=1{.}24\times10^{-3}$ \cite{QuiSha87}.  
The quantity $t_e\equiv\kappa t_{r,i}\approx330t_{r,i}$ is the evolution time of the cluster. Note that in the pure evaporative model $t_e\sim40t_{r,i}$, see e.\,g, \cite{Dok91}. Therefore, the cluster contracted with time, and the rate of (neutron) stars collisions should increase steadily until the relativistic stage and final relativistic gravitational collapse.

Additionally, the dissipative effects due to direct stellar collisions increase the contraction rate of the core. Finally, the avalanche contraction of the cluster is possible  \cite{ZelPod65,Dok91,ShaTeu86,QuiSha87}. The estimation \cite{DokEro11} for the collision rate at this stage of evolution predicts a few collisions of neutron stars per day. An estimation for the number of collapses of the galactic nuclei in the observed part of the Universe gives  
\begin{equation}
\dot N_h\sim\frac{4\pi}{3}(ct_0)^3n_gt_0^{-1}\approx0.1
\left(\frac{n_g}{10^{-2}\mbox{Mpc}^{-3}}\right) \mbox{yr}^{-1},
\label{ratecoll}
\end{equation}
where $n_g$ is a mean number density of galaxies in the universe, and $t_0\simeq10^{10}$~years. The (\ref{ratecoll}) is the rate of the clusters entrance to the dissipative stage. The dissipative stage lasts several years \cite{DokEro11}, so we expect $\sim1$ recurrent sources similar to FRB~121102 in the sky.

\section{Rate of FRBs generation }

The FRBs rate is estimated to be $\sim10^4$ event per day from all sky, which corresponds to $\sim2\cdot10^4$~year$^{-1}$~Gpc$^{-3}$ from the distances $<3.3$~Gpc (red-shifts $z<1$) \cite{Tom13,CalKanWei16}. This is consistent with optimistic scenario of neutron star collision in galactic binary systems, while more realistic scenarios give $\sim10^3$~year$^{-1}$~Gpc$^{-3}$ for the collision rate \cite{Tom13}. The accurate intensity distribution function of FRBs was presented in \cite{Lietal17}. Now we will estimate the possible neutron star merge rate in clusters. 

We consider the case of the neutron star cluster without the central supermassive black hole (the opposite case was considered, e.\,g., in \cite{DokEroOze98}). The close approach of two neutron stars is accompanied by the loose of kinetic energy due to the radiation of gravitational wave, which can be enough for the formation of the gravitationally bound short-lived neutron star binary \cite{QuiSha87}. The cross-section for the formation of the binary is \cite{QuiSha87}
\begin{equation}
\sigma_{\rm cap}\approx\frac{3}{2}\pi r_{g}^{2}{\left(\frac{c}{v}\right)}^{18/7},
 \label{sigmacap}
\end{equation}
where $r_g=2Gm/c^2$, and $m$ is the neutron star mass.
The corresponding rate of the neutron star collisions in the cluster is
\begin{equation}
\dot N_{c}=\frac{1}{2}Nn\sigma_{\rm cap}(\sqrt{2}v)\simeq9\sqrt2{\left(\frac{v}{c}\right)}^{17/7}
\frac{c}{R},
\end{equation}
where $N=M/m$, and $n=3N/(4\pi R^3)$. The mean rate of FRB is
\begin{equation}
\dot n_{c}=\dot N_{c}n_gf_c\simeq 3.4\times10^4\left(\frac{M}{10^6M_\odot}\right)^{17/14}\left(\frac{R}{2\times10^{-4}\mbox{~pc}}\right)^{-31/14}
\left(\frac{n_g}{10^{-2}\mbox{~Mpc$^{-3}$}}\right)\left(\frac{f_c}{10^{-2}}\right)\mbox{~yr$^{-1}$~Gpc$^{-3}$}
\label{ncap1}
\end{equation}
where $f_c$ is the fraction of the galaxies containing the dense neutron stars clusters.
With these normalization factors one has $\sim0.3$ events per cluster per year. So, the (\ref{ncap1}) could easily be at the observable level. 

However the rate of neutron star collisions may be even larger in more dense clusters, because we expect some distribution of clusters over their parameters. In addition to the clusters producing the mean rate (\ref{ncap1}) there are clusters which can produce more frequent (recurrent) bursts similar to FRB~121102. These are the clusters close to the relativistic stage of evolution, described in the Section~\ref{centrsec}, and their emerging rate is given by (\ref{ratecoll}). For these clusters one has 
\begin{equation}
\dot N_{c}\simeq 2.6\times10^{2}{\left(\frac{M}{10^6M_\odot}\right)}^{17/14}\left(\frac{R}{10^{-5}\,{\rm pc}}\right)^{-31/14} \mbox{~yr$^{-1}$},
\label{ncap2}
\end{equation}
and the Schwarzschild radius of this cluster is $2GM/c^2\sim10^{-7}$~pc~$\ll R$. Therefore, the multiple FRBs can also be explained by this model.

After the capture into the pair the resulting orbit of the neutron star is highly elongated. At each successive approach of neutron stars in the binary the interaction of their magnetospheres with the generation of FRBs is possible. Therefore, even for the binary interaction of neutron stars it is possible the generation of additional recurrent FRBs. 

\section{Other observational signatures of neutron stars coalescences}

Let us discuss briefly the possible consequences of neutron stars collisions in the evolved stellar clusters (some of them were discussed earlier in \cite{DokEro11}). In our earlier paper  \cite{DokEro11} the model of the neutron star collisions was appropriated to the description of the avalanche process of the neutron star cluster contraction for explanation of the super-long gamma-ray burst Swift~J164449.3+573451 \cite{Levetal11}. This gamma-ray burst demonstrates the activity during few days. However, nowadays the most popular explanation of this gamma-ray  burst event is a tidal disruption of some star in the vicinity of the central supermassive black hole. 

The neutron star collisions were proposed for explanation of the short gamma-ray burst phenomenon \cite{BNPP84,Pacz86}. The solitary collision of two neutron stars is accompanied by the release of energy $\sim10^{51}-10^{52}$~erg. Probably, some part of gamma-ray bursts is generated in the central stellar clusters of the distant galactic nuclei \cite{DokEroOze98}. It is necessary to explain  why the event FRB~121102 is not associated with the gamma-ray burst. Possibly, it is connected with conditions of gamma-rays collimation in the narrow spatial angle, $\Omega_\gamma\ll1$. It may be that in the direction to the Earth the required jets for gamma-ray bursts were absent, because the probability of registration is diminished in relation $\Omega_\gamma/4\pi$. The recurrent gamma-ray burst may be related with the multiple collisions of neutron stars in the central stellar clusters in the galactic nuclei \cite{DokEroOze98,DokEroOze97}. During the neutron star collisions it is released a lot of gas, which may be supplied for accretion to the central supermassive black hole in the galactic nucleus. In turn, the energy release in accretion is accompanied  by the generation of electromagnetic radiation in the wide range of frequencies. At the same time, the multiple neutron star collisions are the natural physical source for the generation of spherical shocks with sustained energy injection \cite{Dokuch02}.
 
It was supposed also, that neutron stars collisions is accompanied by the generation of relativistic fireballs, which may be gravitationally lensed on the central supermassive black hole, resulting in formation of additional peaks in the gamma-ray radiation \cite{BabDok00}. The characteristic feature of this gravitational lensing is the achromatic flashes of radiation in the different bands of electromagnetic spectrum. Also, the additional peaks in the radiation spectrum arise due to lensing on the stellar mass black holes in the process known as ``mesolensing'' \cite{BarEzo97}.

If the central cluster is surrounded by the gas cloud, then the multiple fireballs are digging out the quasi-stationary rarefied central cavern in the ambient gas. In result, the powerful hidden source of high-energy neutrino may be formed \cite{BerDokHidden,BerDokHidden06}.

The possibility of gravitational wave generation from the coalescence of neutron stars and stellar mass black holes was formulated in the work \cite{QuiSha87}. Gravitational wave astronomy entered recently into the stage of real events observations. Therefore, the model under consideration could be examined in the near future then the sensitivity of observations reach the level of solar-mass compact objects collisions. The conditions in the star cluster are convenient also for the formation of black holes with mass $\sim30M_\odot$, responsible for the generation of the observed gravitational wave signal by the LIGO detector \cite{LigoVirgo}.

\section{Conclusion}

The formation of the dense cluster of neutron stars is almost inevitable consequence of the dynamical evolution of the central stellar cluster in the galactic nucleus. For this reason the multiple neutron star collisions in the galactic nuclei are the very prominent source of the observed FRBs like the recurrent FRB~121102. The multiple events up to several per day could arise from the frequent neutron stars collisions in the clusters. 

The collision of even two neutron stars leads to the formation of close pair with elongated orbit. The orbit shrinks due to the gravitational waves radiation, and the neutron stars approach with one another several times before the final merge. At each approach the magnetospheres of neutron stars overlap and interact. This interactions can lead to the multiple FRBs at the time-scale of minutes during several orbital cycles before the merger \cite{LipPan96}, similar to that was observed in \cite{Schetal16}. Therefore this model predicts the cluster structure of the bursts arrival times for some sources in the form of several pulses per burst. The pulses could be produced either when two neutron stars approach each other at the periastron of highly elongated orbits, or when they finally collide.  
These are actually two different ways of producing FRBs, and they should be somehow different observationally. Multiple FRBs produced in the former case should be weaker, and will not be associated with strong gravitational wave burst. But final FRB produced in the later case might be much stronger, and should be accompanied by a gravitational wave burst. 

The potentially  intriguing observational signature of the recurrent neutron star collisions is  the registration of the corresponding gravitational wave bursts by the laser interferometers LIGO/VIRGO \cite{CalKanWei16} or by the next generation of gravitational wave detectors \cite{DaiWanWu16,Wanetal16,Abbetal-fin}.

\section*{Acknowledgments}

This work was supported in part by the Research Program OFN-17 of the Division of Physical Sciences of the Russian Academy of Sciences.

\end{document}